# Predicting and Forecasting the Price of Constituents and Index of Cryptocurrency Using Machine Learning

Reaz Chowdhury[1], M. Arifur Rahman[2], M. Sohel Rahman[3], M.R.C. Mahdy[1,4*]

[1]*Department of Electrical & Computer Engineering, North South University, Bashundhara, Dhaka 1229, Bangladesh*

[2]*Department of Accounting & Finance, North South University, Bashundhara, Dhaka 1229, Bangladesh*

[3]*Department of Computer Science & Engineering, Bangladesh University of Engineering & Technology, West Palasi, Dhaka 1205, Bangladesh*

[4]*Pi Labs Bangladesh LTD, ARA Bhaban,39, Kazi Nazrul Islam Avenue, Kawran Bazar, Dhaka 1215, Bangladesh*

*Corresponding Author: mahdy.chowdhury@northsouth.edu

**Abstract**

At present, cryptocurrencies have become a global phenomenon in financial sectors as it is one of the most traded financial instruments worldwide. Cryptocurrency is not only one of the most complicated and abstruse fields among financial instruments, but it is also deemed as a perplexing problem in finance due to its high volatility. This paper makes an attempt to apply machine learning techniques on the index and constituents of cryptocurrency with a goal to predict and forecast prices thereof. In particular, the purpose of this paper is to predict and forecast the close (closing) price of the cryptocurrency index 30 and nine constituents of cryptocurrencies using machine learning algorithms and models so that, it becomes easier for people to trade these currencies. We have used several machine learning techniques and algorithms and compared the models with each other to get the best output. We believe that our work will help reduce the challenges and difficulties faced by people, who invest in cryptocurrencies. Moreover, the obtained results can play a major role in cryptocurrency portfolio management and in observing the fluctuations in the prices of constituents of cryptocurrency market. We have also compared our approach with similar state of the art works from the literature, where machine learning approaches are considered for predicting and forecasting the prices of these currencies. In the sequel, we have found that our best approach presents better and competitive results than the best works from the literature thereby advancing the state of the art. Using such prediction and forecasting methods, people can easily understand the trend and it would be even easier for them to trade in a difficult and challenging financial instrument like cryptocurrency.

## 1. Introduction

Cryptocurrencies are a subset of virtual currencies that use cryptography for security. These are decentralized and open source currencies and hence function on a peer-to-peer basis. Cryptocurrencies mostly use a very complex cryptographic algorithm, that requires connected network of computers to conduct computationally expensive mathematical operations [1]. Cryptocurrencies have a built-in implementation of cryptography in their design. At present, people are using cryptocurrencies to implement a new form of economy, because of its cheapness, online, and anonymous means of exchange. A list of cryptocurrencies and their prices can be found at https://coinmarketcap.com, which lists more than 2175 cryptocurrencies of varying types. Cryptocurrencies feature certain computer protocols that are out of any government control. These currencies are unregulated and highly volatile [1]. As a result, it can quickly devalue overnight. These currencies have aggressive swings in their prices, as it is largely based on public perception. It is therefore very hard to make related risk assessment at any moment. With the increase of the prices of cryptocurrencies, mining has also turned into a very advantageous business for the people [2]

One of the most valuable and decentralized cryptocurrencies is Bitcoin, which was introduced by Satoshi Nakamoto on October 31, 2008 [3]. It captures around 35% of the total market capitalization [4]. Bitcoin's greatest innovation is blockchain, which was introduced to solve the issue of double spending as well as to disrupt the control of centralized parties in the transaction of values. The blockchain is the technology in which a record of any financial and economic transactions made in any cryptocurrency are maintained using cluster of computers that are linked in a peer-to-peer network. In simple terms, it is a powerful technology, which has the capacity to maintain permanent records of commercial transactions, transfer of assets and contracts, financial records, and intellectual property [5]. Blockchain is completely a public ledger that is made up of blocks and any node connected on the Bitcoin network can process and clear a transaction by posting the transaction [6].

While the prices of cryptocurrencies have gone up since 2016 with great fluctuation, the enthusiasm of people to invest more and more in these virtual currencies stays more or less constant. These virtual currencies are nowadays used in official cash flows and exchange of goods as a result, in recent years, various physical approaches and modelling techniques have been introduced by researchers and scholars to model the price of cryptocurrencies and to analyze the spontaneity of the market for making real decision support systems [4,7]. These techniques include, but are not limited to, various dynamic topic modelling, machine learning, data mining, and text mining approaches. Moreover, to study the cryptocurrency market, agent based artificial financial market and genetic programming for finding attractive technical patterns have also been proposed [8,9]. In addition, as cryptocurrencies are correlated [10], the cross correlation between price changes of various cryptocurrencies using random matrix theory and minimum spanning tress have also been studied [11]. In recent years, different machine learning algorithms and techniques had also been taken into account to generate abnormal profits by exploiting the inefficiency of the cryptocurrency market [12]. The involvement of some cryptocurrencies including Bitcoin in illicit activities can also be accurately measured using machine leaning approach [13].

The aim of this paper is to predict and forecast the close price of cci30 and constituents of cryptocurrencies thereby helping in minimize the risk in cryptocurrency market. We have collected and analyzed the historical data from https://coinmarketcap.com and have applied various machine learning approaches, employing software RapidMiner [14], to find and analyze the close price of nine cryptocurrencies and for the index, cci30. In particular, we have considered an ensemble learning method, gradient boosted trees model, neural net model, K-Nearest Neighbor (K-NN) model, and have analyzed the performance thereof using the standard measures/metrics from the literature. As the volatility of these currencies is extremely high, it may not be possible to model these currencies using just one algorithm for either prediction or forecast; this actually motivates us to consider and study four different models in this paper. Using these models, we can easily observe the behavior of these currencies and decide which algorithms could be better for prediction and forecasting of the close price thereof.

## 2. Related Works

Cryptocurrency is a new digital asset in finance, which has extremely high volatility as compared to almost all other financial instruments. As a result of its high volatility and price fluctuations, a very limited number of articles, to the best of our knowledge, exists in the literature that deal with predicting the price fluctuations.

Xiaolei et al. [15] have proposed three models; SVM, RF model and Light Gradient Boosting Machine to forecast the price trend of the cryptocurrency market, where the daily data of 42 kinds of cryptocurrencies have been combined with key economic indicators. Light Gradient Boosting Machine stands out to be a better model compared to other models. In case of forecasting performance in the first category of training set, the test set is the true subset of the training set. In this case, the accuracy obtained using Light Gradient Boosting Machine is 0.776 for 2 months, 0.881 for 2 weeks, 0.762 for 2 days and for the first day of the period, the accuracy is 0.7622 for 2 months, 0.905 for 2 weeks and 0.548 for 2 days. In case of forecasting performance in the second category of training set, the test set is not a subset of the training set. In this case, the obtained accuracy using Light Gradient Boosting Machine is 0.607 for 2 weeks, 0.476 for 2 days and for the first day of the period, the accuracy is 0.952 for 2 weeks and 0.93 for 2 days.

Lahmiri et al. [16] have proposed two deep learning techniques, namely, deep learning neural network (DLNN) and generalized regression neural networks (GRNN) to forecast the price of Bitcoin, Digital Cash and Ripple. In case of Bitcoin, RMSE obtained using DLNN and GRNN are $2.75 \times 10^3$ and $8.80 \times 10^3$, respectively. In case of Digital Cash, these values are 19.2923 and 50.2418 and in case of Ripple, 0.0499 and 0.3115. It is seen that, in case of Bitcoin and Digital Cash, the RMSE value obtained using DLNN and GRNN are extremely high, whereas in case of Ripple, this value is comparatively lower than others.

Kim et al. [17] have analyzed user comments in online cryptocurrency communities to predict fluctuation in the prices of the cryptocurrency and in the number of transactions thereof. The accuracy achieved for the predicted fluctuation in Bitcoin price and in Bitcoin transaction are 50.538% and 48.387% for 13 days. For Ethereum, the accuracy of price fluctuation and transaction fluctuation are 49.425% and 51.149% for 13 days. In case of Ripple, the accuracy of Ripple price fluctuation is 63.200 for 13 days. Notably, Ripple transaction fluctuations have not been considered by Kim et al.

Greaves et al. [18] have proposed transaction graph data by collecting Bitcoin transactions to predict the Bitcoin prices. They have used four classification models: baseline, logistic regression, SVM and neural network model. The accuracy obtained using these models are 53.4% for baseline, 54.3% for logistic regression, 53.7% for SVM and 55.1% for neural network respectively.

Barnwal et al. [19] have proposed an approach of stacking with neural network for cryptocurrency investment. Bitcoin data has been used and obtained from quandl to predict the direction of Bitcoin's price. Extreme gradient boosting, K-NN, Light Gradient Boosting Machine are used as discriminative classifiers to create stacks, which are optimized over one layer of neural network to model the direction of price of the cryptocurrencies. Two time period are considered. Apr-May 2018 time period has been used to create level-1 data in stacking, and from June-July 2018, stacked generalizer's performance is compared to the rest of the models. Accuracies of extreme gradient boosting, SVM, K-NN, light gradient boosting, Random Forest, Logistic Elastic Net, Naïve Bayes, Linear Discriminate Analysis, Quadratic Discriminate Analysis and stacked Generalization for April-May (June-July) period have been found to be 0.57 (0.46), 0.48 (0.50),

0.59 (0.52), 0.61 (0.52), 0.55 (0.50), 0.52 (0.50), 0.52 (0.50), 0.54 (0.48), 0.55 (0.52), 0.52 (0.54) respectively.

McNanny et al. [20] have used recurrent neural network (RNN), long short time memory (LSTM) network and ARIMA model to predict the direction of Bitcoin price in USD. RNN and LSTM are two deep learning pipelines, that outperformed the ARIMA forecasting model. Root mean squared error (RMSE) is used to evaluate and compare the regression accuracy and an 80/20 holdout validation strategy is used to instrument the validation of models. As a result, the accuracy and RMSE obtained using ARIMA model are 50.05% and 53.74%. Using RNN (LSTM) model, the accuracy and RMSE obtained are 50.25% (52.78%) and 5.45% (6.87%).

Bakar et al. [21] have implemented an ARIMA forecasting method to determine the accuracy of Bitcoin exchange rates in a high volatility environment. The have achieved absolute percentage errors of 1.4% and 9.3% for September 2017 and October 2017 respectively with a mean absolute percentage error between forecast and actual value is 5.36%.

Rebane et al. [22] have presented an ARIMA model and seq2seq recurrent deep multi-layer neural network (sq2seq) utilizing a varied selection of input types. Cumulative error obtained using seq2seqA, seq2seq B, seq2seq c and Arima models are 1.00, 0.89, 0.45, 1.73 respectively. They have also presented visual comparison of Bitcoin prediction over 40 days, where the results of seq2seq A, seq2seq B, seq2seq c and Arima models are mostly seen to be extremely deviated from the true values.

## 3. Materials and Methods

### 3.1 Dataset

Our dataset includes seven-day week daily data which we have obtained from https:\\coinmarketcap.com. The constituents and predictive models that we have considered are given in Table 1. We have considered seven attributes [see Table 2] and divided our data into two subsets- testing and training. By creating different models in RapidMiner, we have predicted the close price of the constituents and cci30 for the month January 2019 based on historical data. In order to perform gradient boosted tree model and neural network model, we have divided our data of constituents of cryptocurrencies into two subsets- training and testing datasets [see Table 2]. In case of ensemble learning method, the dataset is given in Table 4. For K-NN model, the dataset contains only the training phase for all cryptocurrencies as shown in Table 3 with no testing phase because we plan to forecast the values of the month of January 2019.

### 3.2 Performance Metrics

The performance metrics we have used in this paper are root mean squared error (RMSE), prediction trend accuracy, absolute error, relative error, squared error, correlation, and squared correlation, which are achieved through "Performance (Regression)" and "Forecasting Performance" operators.

Root mean squared error (RMSE) is the measure of the differences between values predicted by the model and the actual values. Prediction trend accuracy measures the average of times a regression prediction was able to correctly predict the trend of the regression. Absolute error is the average absolute deviation of the prediction from the actual value, where the values of the label attribute are the actual values. Relative error is the average of the absolute deviation of the prediction from the actual value divided by actual value.

Squared error chooses the model with the smallest average squared error value. Correlation returns the correlation coefficient between the "label" and "prediction" attributes. Squared correlation returns the squared correlation coefficient between the "label" and "prediction" attributes.

### 3.3 Methodology

Machine learning approaches can play a wide range of critical roles in the finance domain especially when it comes to predict the prices of financial instruments in general and cryptocurrencies in particular. Starting from managing cryptocurrency portfolio [23] to predicting the fluctuations in the prices of cryptocurrencies transactions [17], machine learning stands out to be one of the best approaches and techniques. Machine learning techniques can be integrated into business intelligence systems for making real life decisions [24]. Effort to predict and analyze the price of cryptocurrencies have been a very challenging one due to its high volatility and price fluctuations. Thus, it is hoped that, ML techniques will bring a new dimension in this domain and as it has already been discussed in the literature review, we already have a number of ML based approaches in this domain in the literature.

In this paper, we have used the widely used software called RapidMiner as it supports all steps of a data mining process [12]. In RapidMiner software, for performing data analysis usually graphs, plots, charts and tables are used in which one can easily visualize the output and also compare between one or more attributes and models. For a machine to predict and forecast the future close price of any cryptocurrency, it is essential to train the machine to learn from the given dataset. Based on these datasets, models will be created applying different algorithms and thus the prediction/forecasting task will be accomplished [25].

#### 3.3.1 Predictive Model: Gradient Boosted Trees

Gradient boosted trees model is very advantageous especially in the context of price prediction for a number of reasons as follows. Firstly, it is not required to normalize the data in this case as it is sensitive to arithmetic range of data and features. Secondly, it is a very scalable machine learning model due to its construction process and finally, it is also a rule-based learning method [26]. A number of works dealing with prediction and forecasting of sales as well as cryptocurrency prices in the literature have successfully employed gradient boosted trees model [12,15,27].

Gradient boosted trees model is an ensemble of either regression or classification tree models, which is a forward learning ensemble method that obtains predictive results through gradually improved estimations. For predicting the close price of cryptocurrency of the month January 2019, we have considered the attributes of dataset stated in Table 2 and provided the historical price of the constituents as shown in Table 3. We have further used RapidMiner to optimize the parameters of the model, which runs through various permutations to get the best value of its parameters. The parameter, number of trees is tuned to 500 through optimization, while the other parameters are set as default. The "Performance (Regression)" and "Forecasting Performance" operators are used to determine the performance of our testing dataset. Chart 1(a) illustrates the performance of gradient boosted trees model using all the metrics mentioned in Section 3.2 for all nine constituents; the actual values have been provided in tabular format in supplement S1. The model I given in Figure 1. Finally, to visualize the comparison between original close price and predicted close price of all nine constituents of the month January 2019, we have created graphs which are shown in figure 2 (a)-(i). A small part of leaf of gradient boosted tree model is shown in supplement S2.

### 3.3.2 Predictive Model: Neural Net

An artificial neural network, also called a neural network, is a mathematical and computational model that is inspired by the structure and functional aspects of biological neural networks. A neural network consists of an interconnected group of artificial neurons, and it processes information using a connectionist approach to computation. Neural networks can be employed to model complex relationships between inputs and outputs or to find patterns in data to predict the price of cryptocurrencies [28,29].

We have considered the same model used in [25], the outlook of which is given in figure 3. We have given the historical prices of all seven attributes as input in the training dataset. In the testing dataset, we have provided the attributes of the month January 2019. Set role operator is used to set the attribute name as "Date" and target role as "id". A windowing operator is used, which will create examples from the value series data set by windowing the input data we have provided. The parameters of this operator are changed. The parameters "series representation" is used to represent the series values and it is set as "encode_series_by_examples". The parameter "window size" is the width of the used window and the "step size" is the distance between the first values. Both of these parameters are set to 1. Create single attribute and create label options are checked as we have the close price as our label and we are predicting the future close prices of the cryptocurrencies depending on the close prices of the past. On the other hand, in testing side no label attribute is created on another windowing operator as it will be predicted by the model. Horizon is the distance between the last window value and the value to predict. A sliding window validation operator is used to enclose sliding windows of training and tests in order to estimate the performance of a prediction operator. The parameters of these operator are also changed. The parameter "training window width" is the number of examples in the window, which is used for training and it is tuned to 4. The "training window step size" is the number of examples the window is moved after each iteration and it is tuned to 1. The "test window width" is the number of examples, which are used for testing and it is tuned to 4. Horizon is the increment from last training to first testing example and it is kept at default state i.e. 1. The sliding window validation operator is a subprocess, so it has two phases- training and testing. The training phase have been provided with a neural net model. The parameters of this operator are training cycles, learning rate and momentum. We have trained the model with 500 training cycles. Learning rate determines the change of weights at each step and it is tuned to 0.03. The momentum adds a fraction of the previous weight update to the current one as a result, it avoids local maxima and smoothen the optimization directions; it is tuned to 0.9. The testing side has a "Apply Model" operator with a "Forecasting Performance" or "Performance (Regression)" operator. Chart 1(b) illustrates the performance of neural net model using all the metrics mentioned in Section 3.2 for all nine constituents; the actual values have been provided in tabular format in supplement S3. To visualize the difference between original and predicted close price of the constituents, we have shown comparative graphs in figure 4 (a)-(i).

### 3.3.3 Predictive Model: Ensemble Learning Method

In an ensemble learning method multiple machine learning algorithms or learners are strategically generated and combined together in order to solve one particular computational intelligence problem. Using ensemble learning method, we can construct set of models and combine them through voting process, as a result of which, the likelihood of an unfortunate selection of a bad model for prediction can be reduced. This method helps to overcome the biases and error rates of the individual (weak) models by combining them through creating a strong learner by uniting some weak learners. The training dataset plays the role of most effective contributor to the error in the model so we have taken a large training dataset with seven attributes and close price as the label attribute. In recent years, several ensemble learning techniques have been proposed to find and predict the prices of cryptocurrencies [12,30].

In this paper, we have created an ensemble learning method in order to get better results by employing different models together, which is shown in figure 5. As the "Split Data" operator produces the desired number of subsets of the given dataset so it is used to partition our data into subsets. The parameters of this operator are "partitions" and "sampling type". The "partitions" parameter is used to split our data for training and testing in the ratio of 0.6 and 0.4. The "sampling type" parameter is set as "linear sampling" to simply divide our dataset into partitions without changing the order of the examples. The "Vote" operator is a subprocess, which uses a majority vote for classification or the average for regression. We have given gradient boosted trees, neural net and relative regression operator inside the "Vote" operator. In case of gradient boosted trees model, the number of trees parameter is set to 500 and other parameters are kept at default state. For neural net model, there are 500 training cycles, with the learning rate set to 0.3 and momentum set to 0.2. The relative regression operator learns a regression model for predictions relative to another attribute value. It is a meta regression learner and useful for time series predictions of dataset with large trends. Inside this learner, we have chosen a "Linear Regression" model and the parameters of linear regression are kept at default state. It is seen that, using multiple relative regression learner, it is possible to minimize the error and thus the desired output can be obtained through a voting process. Finally, an "Apply Model" operator is used to apply the models through the voting process with a "Performance (Regression)" operator or a "Forecasting Performance" operator to find the performance vectors/metrics. Chart 1(c) illustrates the performance of ensemble learning method using all the metrics mentioned in Section 3.2 for all nine constituents; the actual values have been provided in tabular format in supplement S4. To visualize the difference between original and predicted close price of the constituents, we have shown comparative graphs in figure 6 (a)-(i).

### 3.3.4 Predictive Model: K-NN

The K-NN (k-Nearest Neighbor) algorithm is a non-parametric method that is based on comparing an unknown dataset with the K training examples, which are the nearest neighbors of the unknown example. This algorithm is basically used for generating a K-Nearest Neighbor model that can be used for either classification or regression. So, it can be easily understood that, K-NN algorithm can play a very important role in forecasting the price of constituents and of cryptocurrencies. Moreover, we have seen its use in case of detecting covert cryptocurrency mining operations, which causes great losses to the organizations [31]. Besides, reverse K-NN method is also used for managing location data of IoT service providers and users based on blockchain with smart contract [32].

In this paper, we have used a model similar to a forecasting model presented in [33]. For forecasting the close price of these currencies using K-NN model, we have taken into account, the attributes "Date" and "Close Price" only. The model is shown in figure 7. We have used our model to forecast the close prices of the month of January 2019. In order to evaluate a time series model, a residual analysis has to be performed. It indicates the difference between our chosen forecasting method and actuals and its determination helps to observe how well the chosen forecast method fits to our historical data that we have provided. We know that, a good forecasting method will yield residuals which are uncorrelated and having mean near to zero. If the mean is not near to zero then, that means that the forecasts are biased. However, cryptocurrencies are correlated [10]. The list of residuals (average) of all cryptocurrencies and index is shown in Table 5. Chart 1(d) illustrates the performance of K-NN model using all the metrics mentioned in Section 3.2 for all nine constituents; the actual values have been provided in tabular format in supplement S5. Figure 8 (a)-(i) illustrates the differences between the original and forecasted close prices of the month January 2019. Figure 10 compares the predicted and forecasted values by all models.

### 3.4 Index

In this paper, we have taken into account the "Cryptocurrencies Index 30 (cci30)", for prediction and forecasting from https://cci30.com. It is a rules-based index designed to objectively measure the overall growth, daily and long-term movement of the blockchain sector [34]. It works by tracking thirty largest cryptocurrencies market capitalization. It was launched on January 1st 2017 and it's starting value was arbitrarily set at 100 on January 1st 2015.

We have downloaded the OHLCV daily values of the index from [34]. We have predicted the daily close price of the index of the month of January 2019 using gradient boosted trees model, ensemble learning method, neural network model and for forecasting the daily close price of the month of January 2019, we have used K-NN algorithm in RapidMiner platform. The training and testing datasets are given in Table 6 for prediction, while we have used only training phase for K-NN model and forecasted the close prices of the month of January 2019. For predicting the close price of the index of the month January 2019 using RapidMiner, we have considered the attributes, which are given in Table 5, except volume and market capital. For forecasting we have used only the "Date" and "Close Price" attributes. The performance vectors of the models are shown in supplement S6 and the comparison graphs are shown in figure 9(a)-(d), which illustrates the comparison between original and predicted close price of the month January 2019 obtained from all models. This comparison is also shown including all models together in figure 10 (j).

### 3.5 Comparison with Previous Results

This section presents a detailed comparison among the results obtained by the four models in this paper and other state of the art methods.

A Gradient Boosting Decision Tree (GBDT) algorithm, Light Gradient Boosting Machine (LightGBM) is adopted in [15] to forecast the price trend of cryptocurrency market, where they have argued that, the robustness of LightGBM model works better than SVM and RF model. Using LightGBM model, maximum forecasting performance in the first category of training sets has accuracy of 0.905 for two weeks and in second category of training sets, the accuracy is 0.952 for two weeks. In Table 7, we have shown a comparison between the results obtained in this paper and that in [15], which shows that our accuracy is 0.924 using ensemble learning method, whereas in [15], the highest accuracy is 0.952 using LightGBM model. However, they have forecasted the price trend (falling, not falling) by combining daily data of 42 kinds of cryptocurrencies with key economy indicators, but we have predicted and forecasted the price of 9 cryptocurrencies and cci30. In paper [15], only six months daily trading data from January 1, 2018 to June 30, 2018 were collected and considered for forecasting, but in our paper, we have taken yearly data, as a result of which our result is slightly less than in [15].

Deep learning techniques have been employed to forecast the price of Bitcoin, Digital Cash and Ripple in [16], where they have demonstrated that long-short term memory neural network (LSTM) performs better than generalized regression neural networks. The RMSE obtained using deep learning LSTM networks for Bitcoin, Digital Cash and Ripple are $2.75 \times 10^3$ 19.2923 and 0.0499. In Table 8, we have shown a comparison between the results obtained in this paper and that in [16], which depicts that the models and learners we have used for prediction and forecast, yielded much better results than the work done in [16]. The RMSE we have obtained by Gradient Boosted Trees for Bitcoin, Doge coin and NEM are 32.863, 0.000 and 0.001.

Non-linear deep learning methods (LSTM and RNN) and ARIMA model are employed in [20] to predict the direction of Bitcoin prices in USD with good accuracy. It has been demonstrated that, LSTM and RNN have performed better than ARIMA model. LSTM gives accuracy and RMSE of 52.78% and 6.87%, while RNN gives accuracy and RMSE of 50.25% and 5.45%; for ARIMA model, the values are 50.05% and 53.74% respectively. In Table 9, we have shown a comparison between the results obtained in this paper and that in [20]. It is seen by comparing that, our maximum accuracy and RMSE are much better than it is in paper [20]. Our highest accuracy and RMSE using gradient boosted trees model are 0.900 and 0.001, and 0.924 and 0.002 using ensemble learning method.

## 4    Conclusion

In this article, we have presented four different models to predict and forecast the close prices of nine constituents and cci30 using machine learning approaches. Our models exhibit a very good performance in overall prediction of the close price of cryptocurrencies, which can be extremely useful for all including public, private, and government organizations as through our models, the trends and patterns of these currencies can be well-understood. However, in case of forecasting, the K-NN model didn't worked very effectively unlike other models, which happened due to the presence of noisy random features and extreme volatility. We have compared our work with the state-of-the-art models from the literature and demonstrated that our models' performance seems better and competitive. We believe and hope that our model will be beneficial for people to observe, understand, and choose their own desired currency from cryptocurrency market.

## List of the Tables and Charts

**Table 1. Names of the constituents and prediction models under consideration in this paper.**

| | |
|---|---|
| Name of the nine Constituents | Bitcoin Cash; Bitcoin; Dash; Doge Coin (DOGE); Ethereum; IOTA (MIOTA); Litecoin; NEM; NEO. |
| Predictive Models and Learners | Gradient Boosted Trees, Neural Net, Ensemble Learning, K-NN |

**Table 2. Attributes of the dataset under consideration.**

| SL. | Attributes of the Dataset | Remarks |
|---|---|---|
| 1. | Date | Date or a trade date is a day at which an order is executed in the market to purchase, sell or otherwise acquire a currency is performed. |
| 2. | Open Price | Open (opening) price is the price at which a currency is first traded on a given trading day. |
| 3. | Close Price | Close (closing) price is the final price at which a currency is traded on a given trading day. |
| 4. | High | High is the highest price at which a currency is traded on a given trading day. |
| 5. | Low | Low is the lowest price at which a currency is traded on a given trading day. |
| 6. | Volume | Volume or volume of trade is the total quantity of contracts traded for a specified currency. |
| 7. | Market Capital | Market capital refers to the total dollar market value of a currency's outstanding contracts. |

**Table 3. Training, testing and Forecasting dataset of the constituents for gradient boosted trees, neural net and K-NN models.**

| Names of Constituents | Training data | Testing and Forecasting data |
|---|---|---|
| Bitcoin Cash | 01.08.2017-31.12.2018 | 01.01.2019-31.01.2019 |
| Bitcoin | 27.12.2013-31.12.2018 | 01.01.2019-31.01.2019 |
| Dash | 14.02.2014-31.12.2018 | 01.01.2019-31.01.2019 |
| Doge Coin (DOGE) | 27.12.2013-31.12.2018 | 01.01.2019-31.01.2019 |
| Ethereum | 07.08.2015-31.12.2018 | 01.01.2019-31.01.2019 |
| IOTA (MIOTA) | 13.06.2017-31.12.2018 | 01.01.2019-31.01.2019 |
| Litecoin | 27.12.2013-31.12.2018 | 01.01.2019-31.01.2019 |
| NEM | 01.04.2015-31.12.2018 | 01.01.2019-31.01.2019 |
| NEO | 25.10.2016-31.12.2018 | 01.01.2019-31.01.2019 |

**Table 4. Training dataset of the companies for ensemble learning method.**

| Names of Constituents | Dataset |
|---|---|
| Bitcoin Cash | 01.08.2017-31.01.2019 |
| Bitcoin | 27.12.2013-31.01.2019 |
| Dash | 14.02.2014-31.01.2019 |
| Doge Coin (DOGE) | 27.12.2013-31.01.2019 |
| Ethereum | 07.08.2015-31.01.2019 |
| IOTA (MIOTA) | 13.06.2017-31.01.2019 |
| Litecoin | 27.12.2013-31.01.2019 |
| NEM | 01.04.2015-31.01.2019 |
| NEO | 25.10.2016-31.01.2019 |

**Table 5. Residuals (average) of all the constituents and index using K-NN model.**

| Names of Constituents | Residuals (average) |
|---|---|
| Bitcoin | -68.28510616239669 |
| Bitcoin cash | -78.30352757348386 |
| Dash | -9.896544294902897 |
| Doge Coin (DOGE) | 3.489083182322858E-5 |
| Ethereum | -26.347865050983437 |
| IOTA (MIOTA) | -0.01377814670751103 |
| Litecoin | 0.968223740614247 |
| NEM | -0.030791453489023235 |
| NEO | -3.602597822081795 |
| Index (cci30) | -186.59374626179212 |

**Table 6. Training, testing and Forecasting dataset of the index for gradient boosted trees, neural net, ensemble learning method and K-NN models.**

| Prediction and Forecasting Models | Training | Testing |
|---|---|---|
| Gradient Boosted Trees | 01.01.2015-31.12.2018 | 01.01.2019-31.01.2019 |
| Neural Net | 01.01.2015-31.12.2018 | 01.01.2019-31.01.2019 |
| Ensemble | 01.01.2015-31.01.2019 | 01.01.2019-31.01.2019 |
| K-NN | 01.01.2015-31.12.2019 | 01.01.2019-31.01.2019 |

**Table 7. Comparison between state-of-the-art model in [15] and our model in this paper.**

| State of the art work | | | | | | | Work of this paper | |
|---|---|---|---|---|---|---|---|---|
| **Forecasting performance in the first category of training sets** | | | | | | | | |
| Forecasting Model | Period (Accuracy) | | | 1st day of the period (accuracy) | | | Model | Accuracy |
| | 2 months | 2 weeks | 2 days | 2 months | 2 weeks | 2 days | Ensemble Learning Method | 0.924 |
| LightGBM model | 0.776 | 0.881 | 0.762 | 0.762 | 0.905 | 0.548 | | |
| **Forecasting performance in the second category of training sets** | | | | 1st day of the period (accuracy) | | | | |
| Forecasting Model | 2 weeks | 2 days | | 2 weeks | 2 days | | | |
| LightGBM model | 0.607 | 0.476 | | 0.952 | 0.93 | | | |

.

**Table 8. Comparison between state-of-the-art model in [16] and our model in this paper.**

| State of the art work | | | Work of this paper | |
|---|---|---|---|---|
| Currency | RMSE | | Currency | RMSE |
| | DLNN | GRNN | | Gradient Boosted Trees |
| Bitcoin | 2.75×10^3 | 8.80×10^3 | Bitcoin | 32.863 |
| Digital Cash | 19.2923 | 50.2418 | Doge Coin | 0.000 |
| Ripple | 0.0499 | 0.3115 | NEM | 0.001 |

**Table 9. Comparison between state-of-the-art model in [20] and our models in this paper.**

| State of the art work | | | Work of this paper | | |
|---|---|---|---|---|---|
| Model | Accuracy | RMSE | Model | Accuracy | RMSE |
| LSTM | 52.78% | 6.87% | Gradient Boosted Trees | 0.900 | 0.001 |
| RNN | 50.25% | 5.45% | Ensemble Learning method | 0.924 | 0.002 |
| ARIMA | 50.05% | 53.74% | | | |

**Chart 1: Performance measurements**

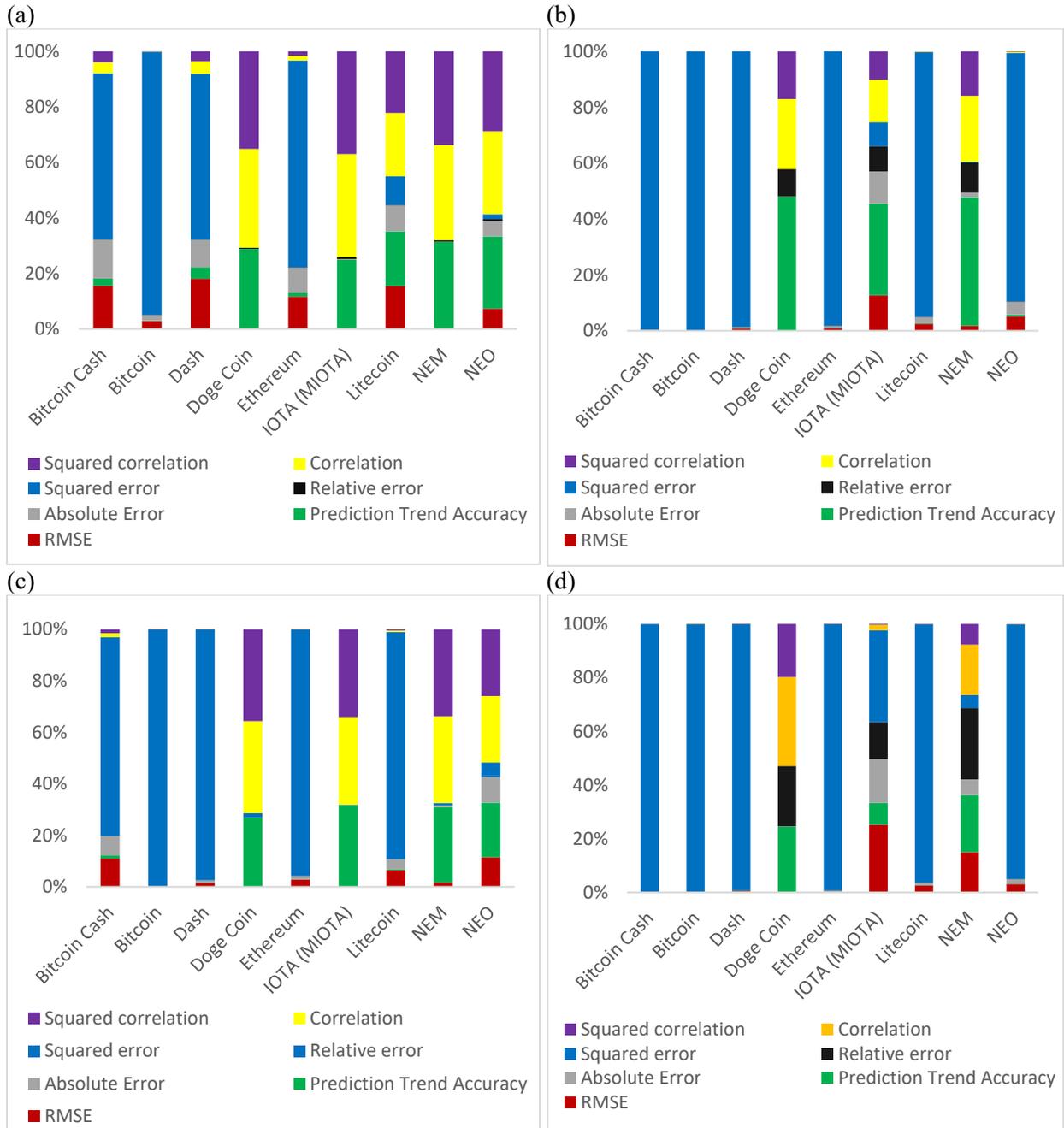

Chart 1: Performance measures of (a) Gradient Boosted Trees model (b) Neural Network model (c) Ensemble learning method (d) K-NN model.

**Figures and Captions List**

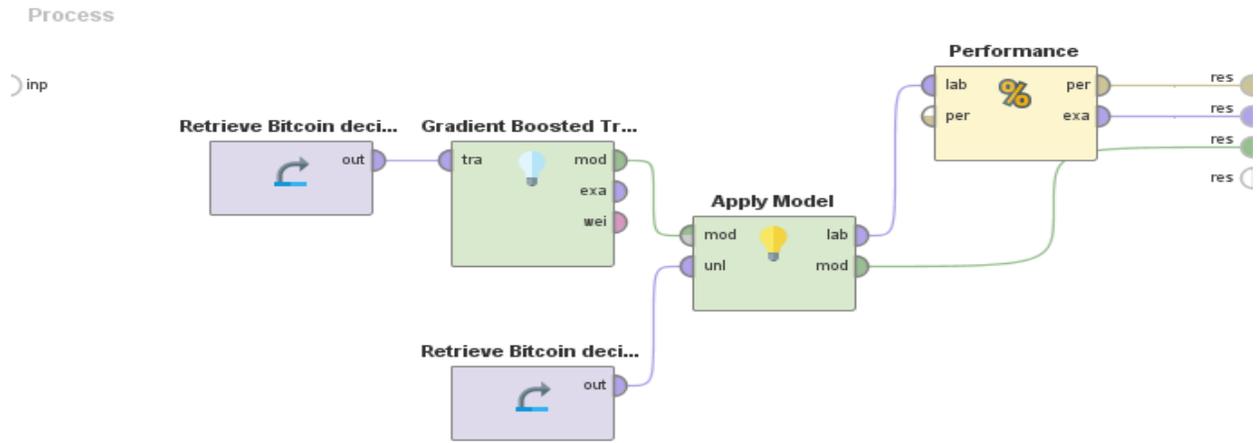

Figure 1. Gradient boosted trees model for predicting constituents and index of cryptocurrencies.

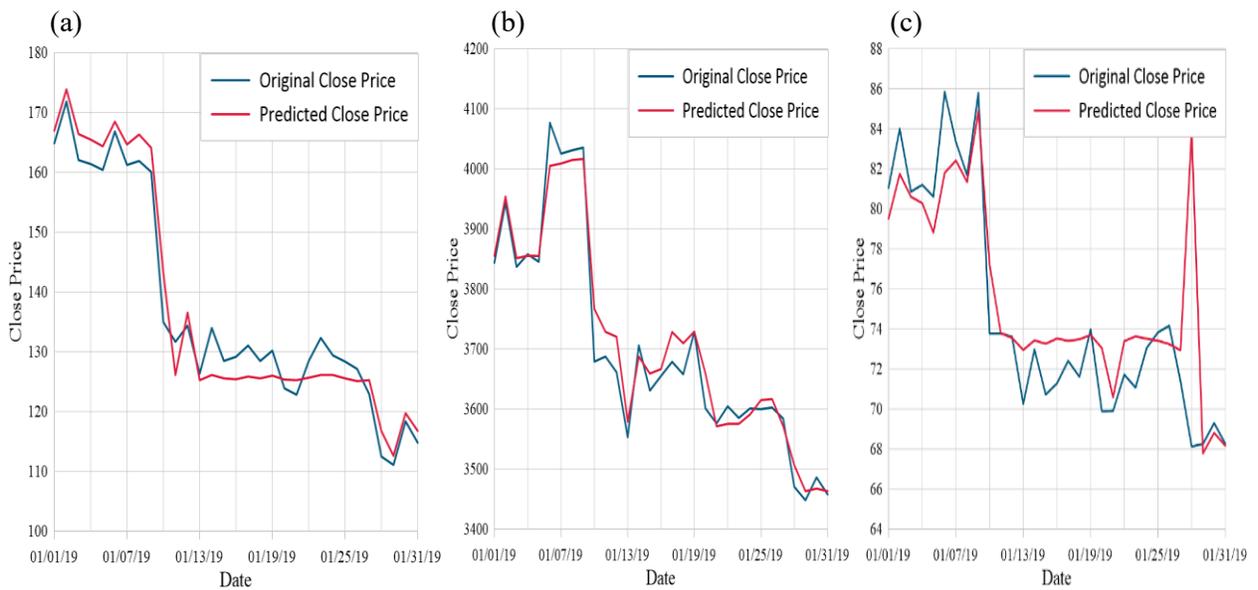

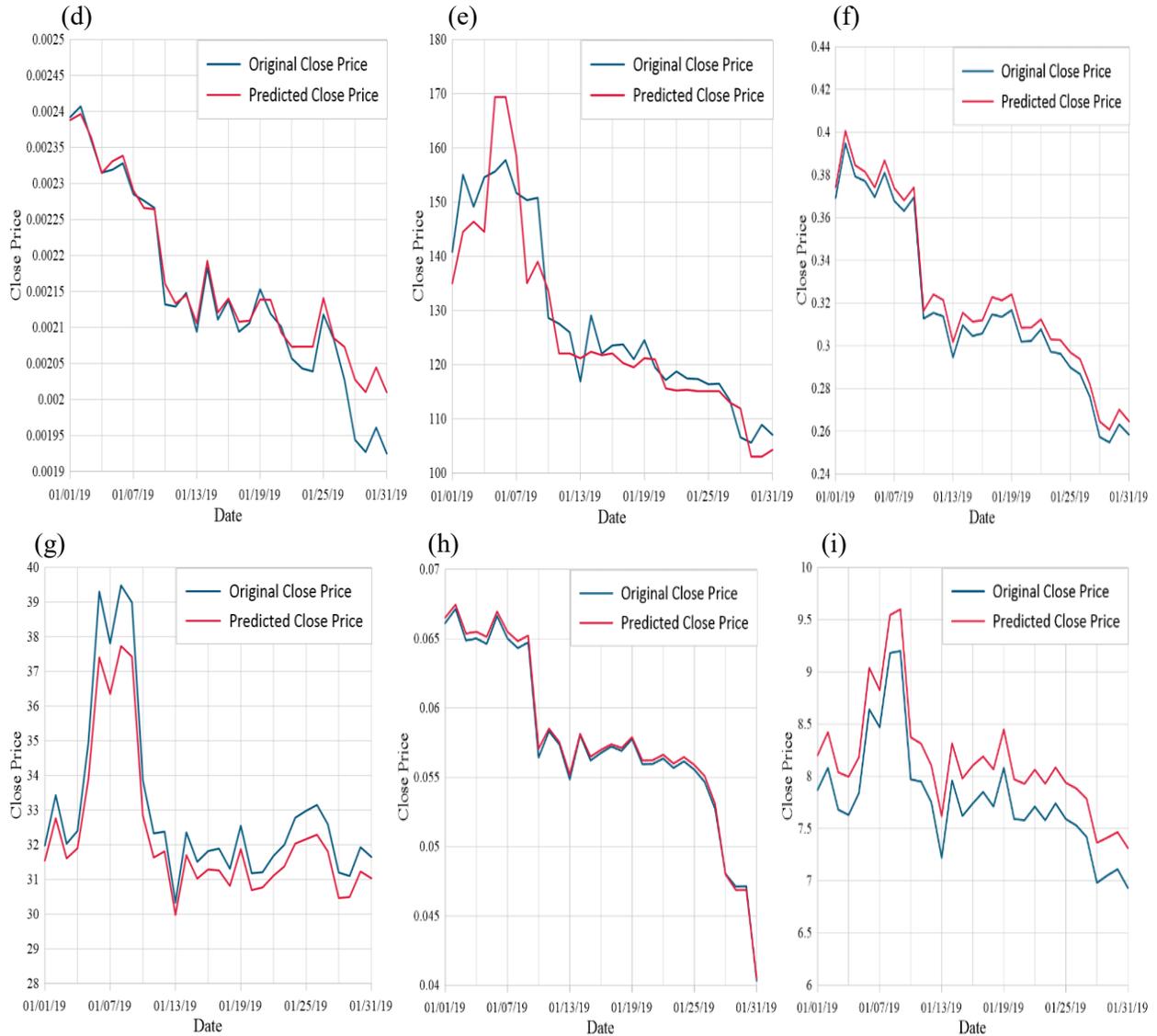

Figure 2. Comparison between original and predicted close price obtained from RapidMiner using model gradient boosted trees for the month January 2019 (a) of constituent Bitcoin Cash. (b) of constituent Bitcoin. (c) of constituent Dash. (d) of constituent Dogecoin (DOGE). (e) of constituent Ethereum. (f) of constituent IOTA (MIOTA). (g) of constituent Litecoin. (h) of constituent NEM. (i) of constituent NEO.

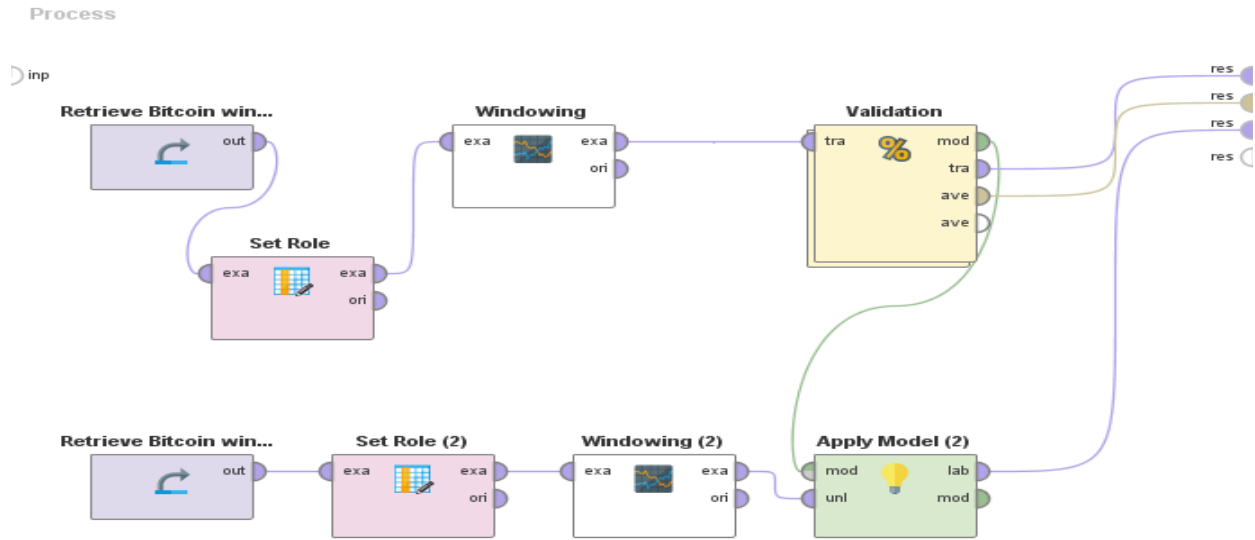

Figure 3. Neural net model for predicting constituents and index of cryptocurrencies.

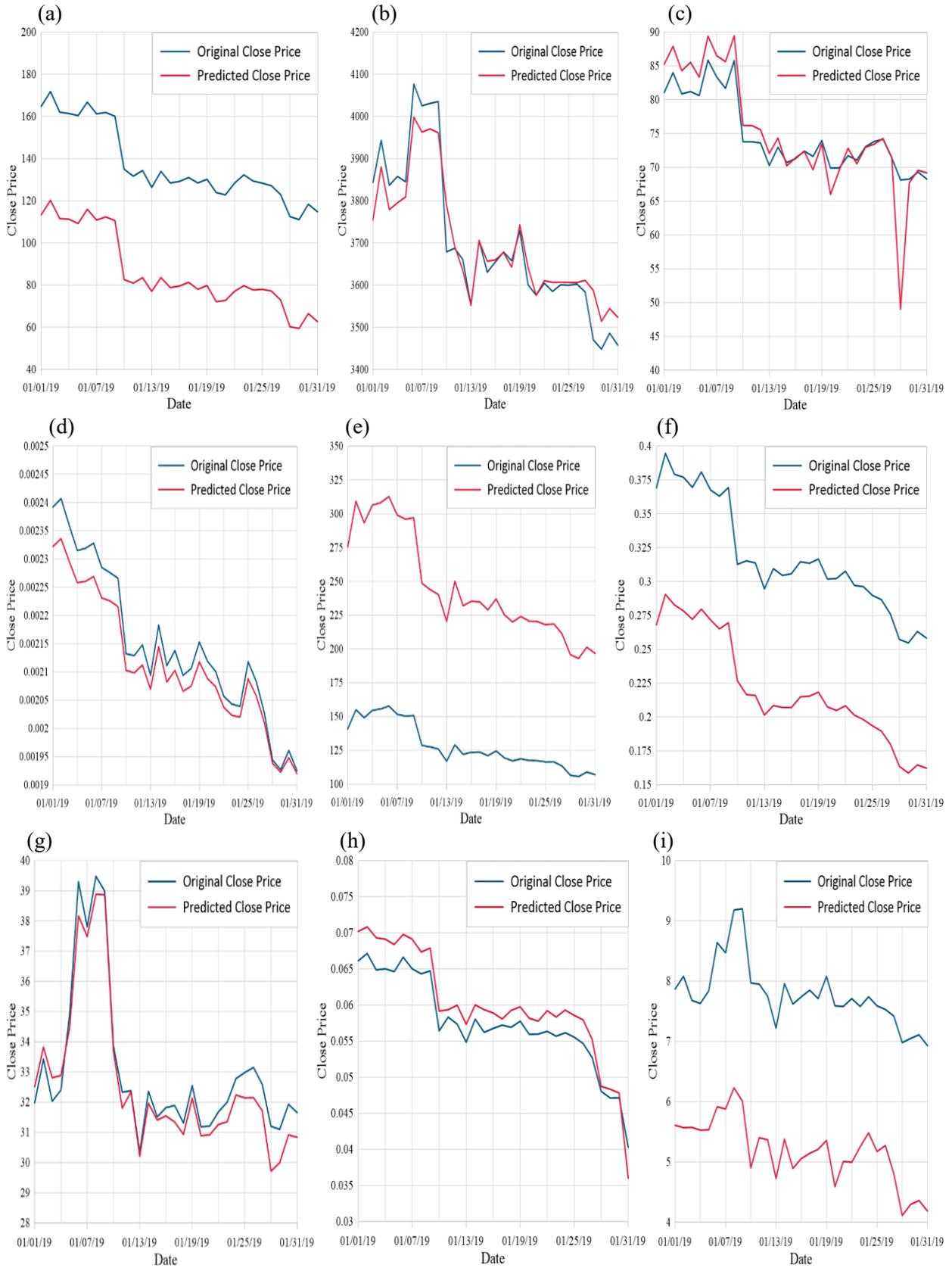

Figure 4. Comparison between original and predicted close price obtained from RapidMiner using neural net model for the month January 2019 (a) of constituent Bitcoin Cash. (b) of constituent Bitcoin. (c) of constituent Dash. (d) of constituent Dogecoin (DOGE). (e) of constituent Ethereum. (f) of constituent IOTA (MIOTA). (g) of constituent Litecoin. (h) of constituent NEM. (i) of constituent NEO.

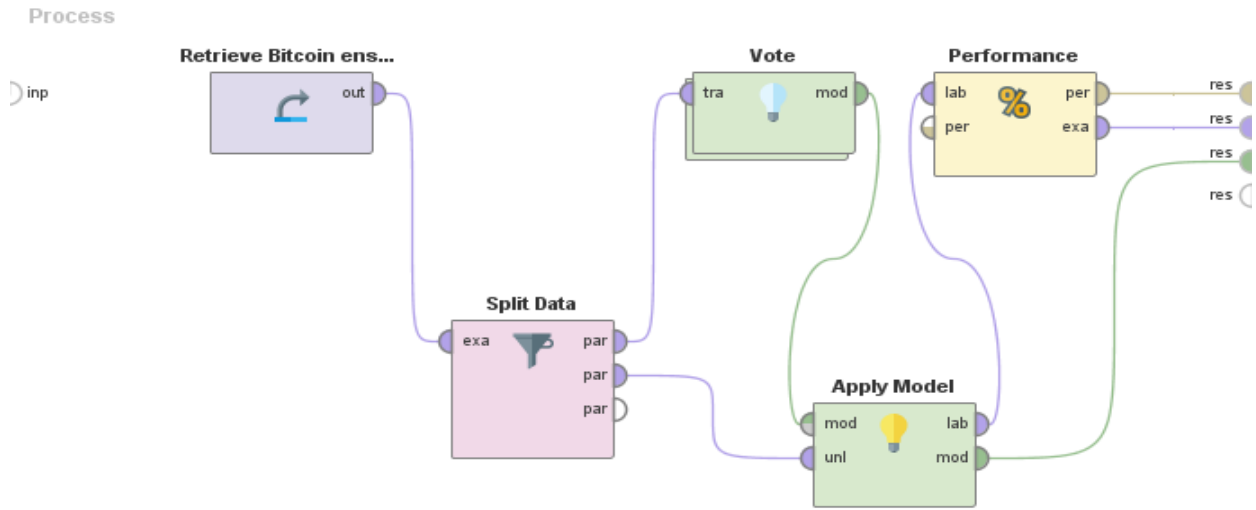

Figure 5. Ensemble learning model for predicting constituents and index of cryptocurrencies.

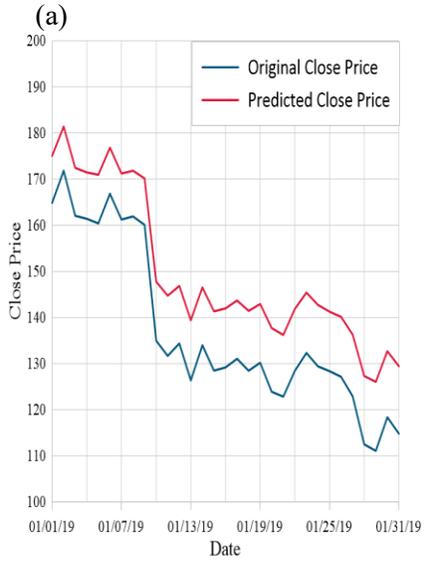
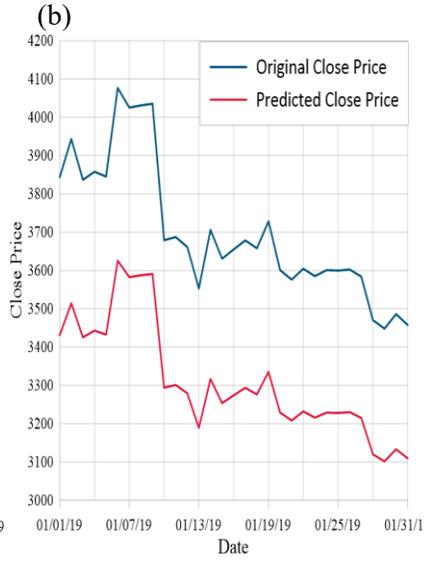
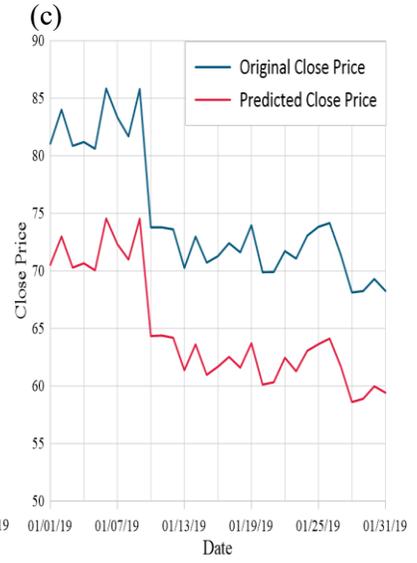
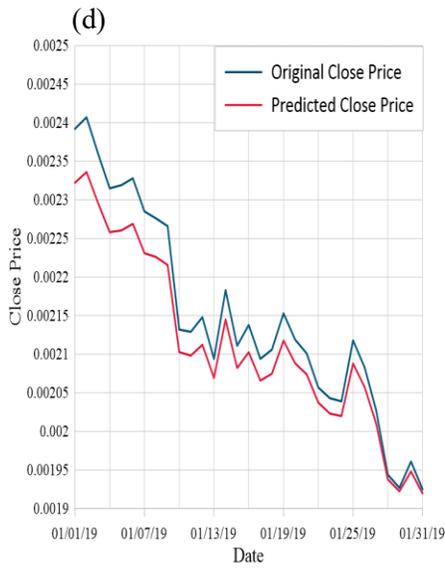
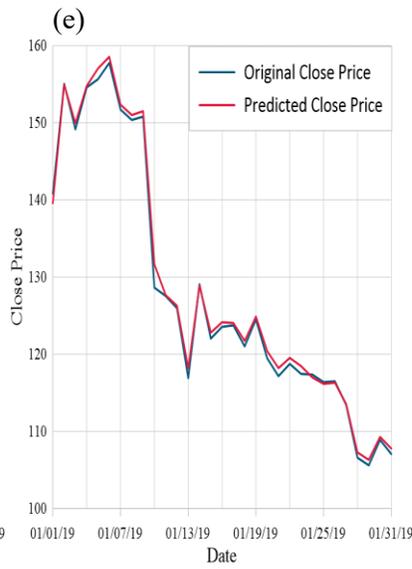
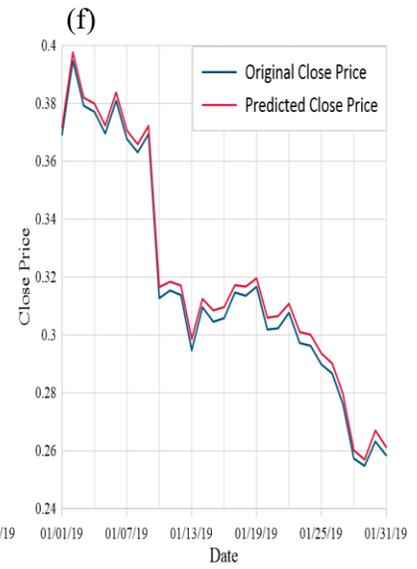
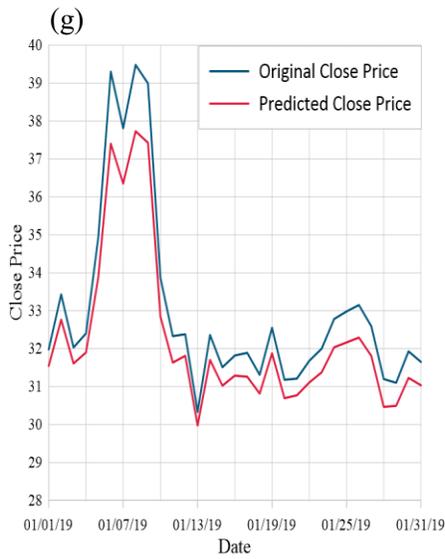
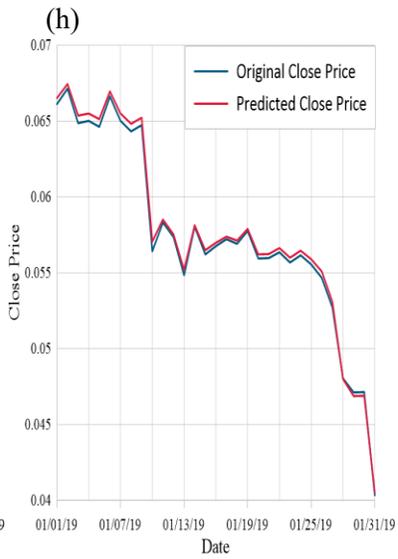
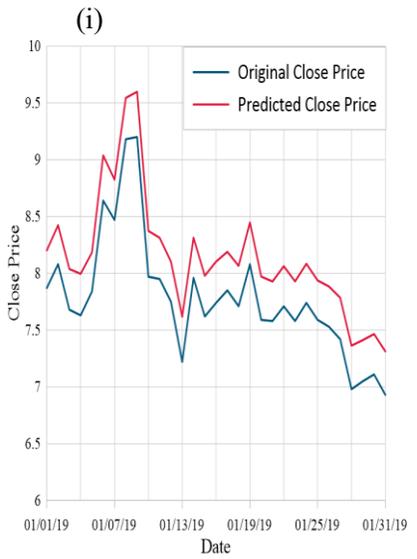

Figure 6. Comparison between original and predicted close price obtained from RapidMiner using ensemble learning method for the month January 2019 (a) of constituent Bitcoin Cash. (b) of constituent Bitcoin. (c) of constituent Dash. (d) of constituent Dogecoin (DOGE). (e) of constituent Ethereum. (f) of constituent IOTA (MIOTA). (g) of constituent Litecoin. (h) of constituent NEM. (i) of constituent NEO.

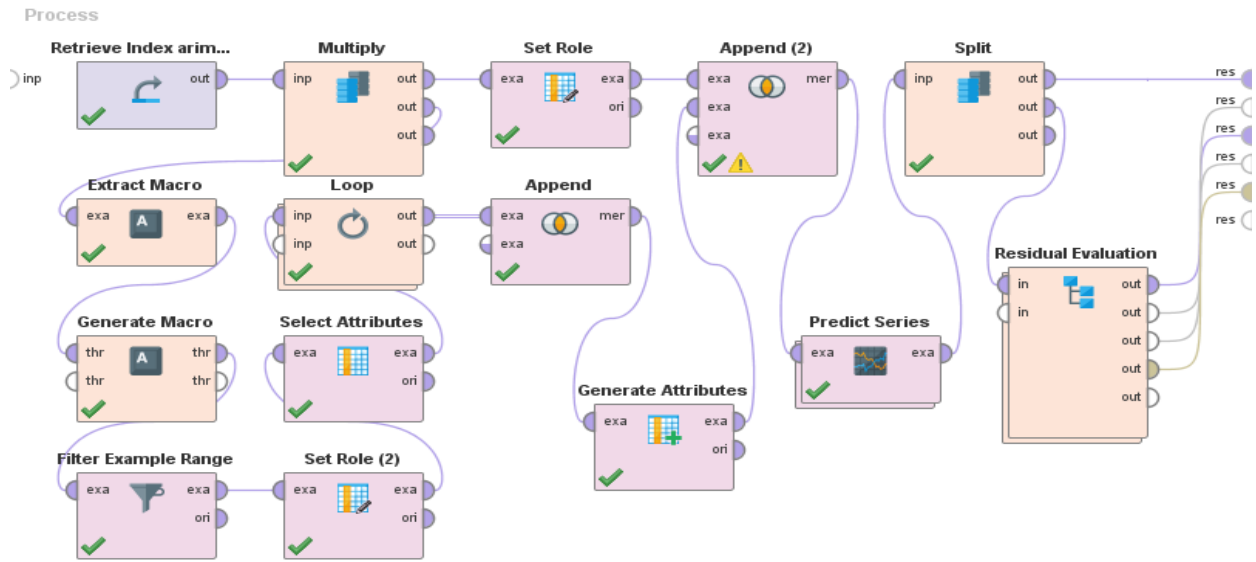

Figure 7. K-NN model for forecasting constituents and index of cryptocurrencies.

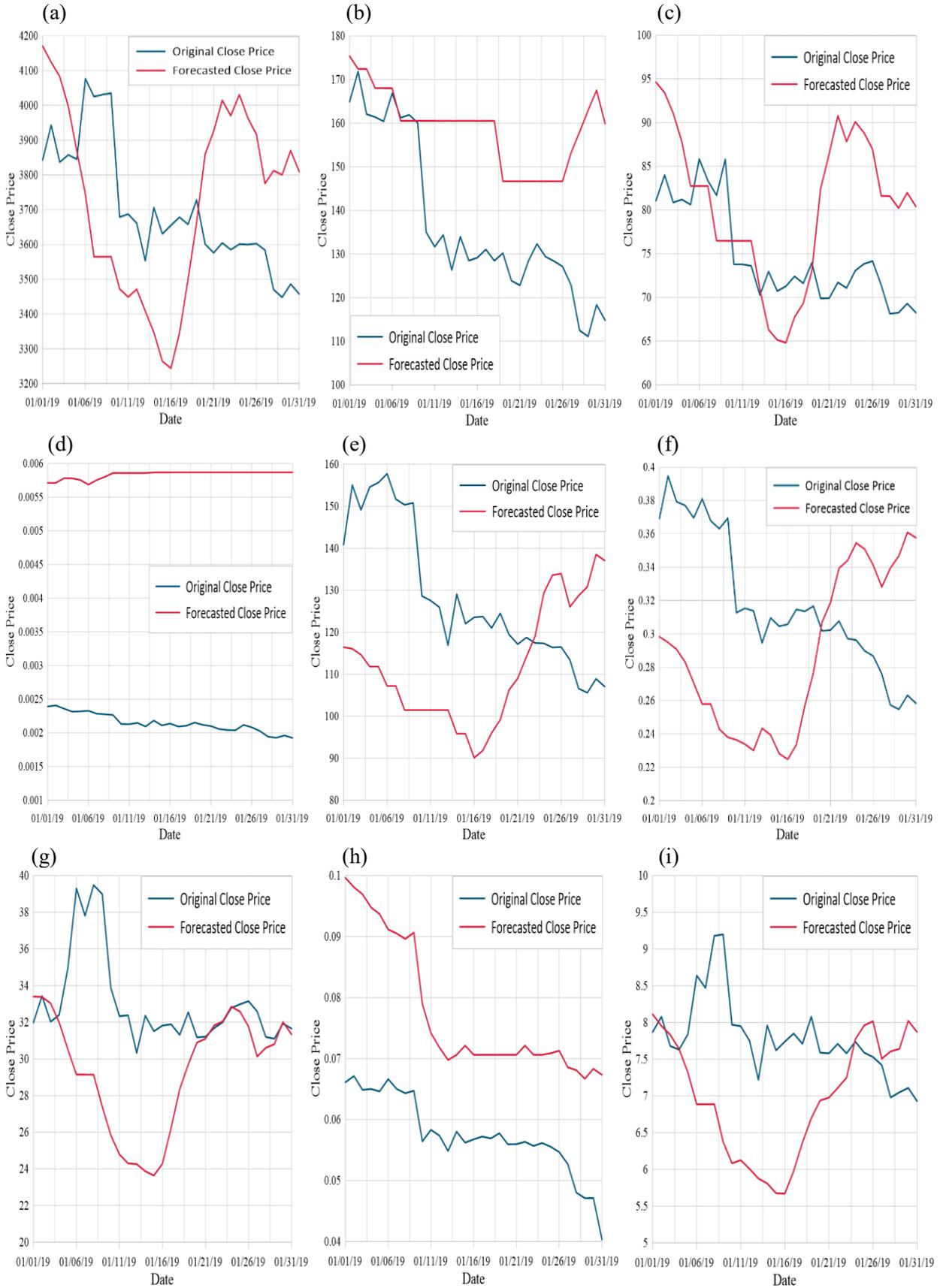

Figure 8. Comparison between original and forecasted close price obtained from RapidMiner using K-NN model (a) of constituent Bitcoin. (b) of constituent Bitcoin Cash. (c) of constituent Dash. (d) of constituent Dogecoin (DOGE). (e) of constituent Ethereum. (f) of constituent IOTA (MIOTA). (g) of constituent Litecoin. (h) of constituent NEM. (i) of constituent NEO.

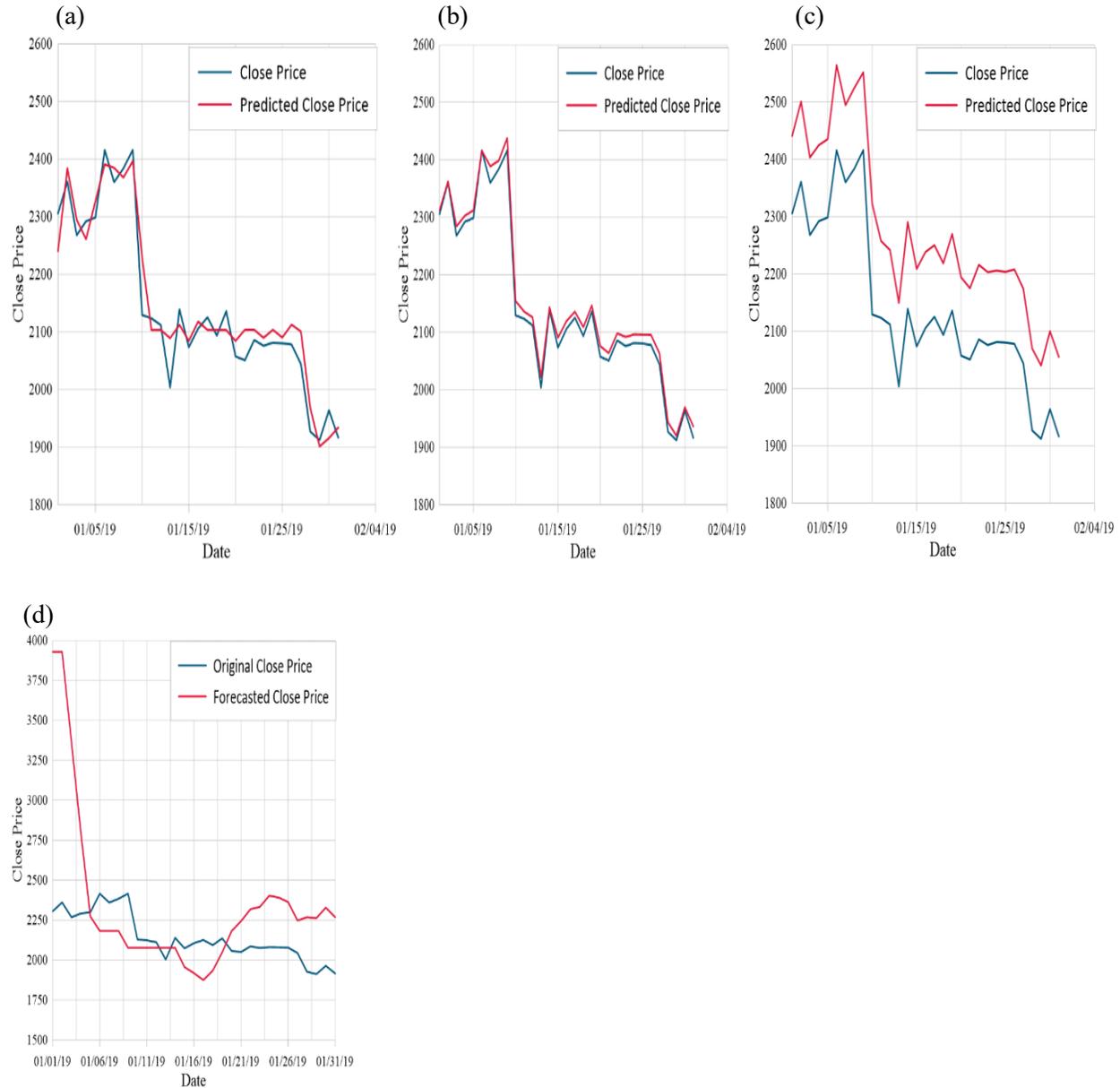

Figure 9. Comparison between original and forecasted close price obtained from RapidMiner of index cci30 (a) using gradient boosted trees model. (b) using ensemble learning method. (c) using neural net model (d) using K-NN model.

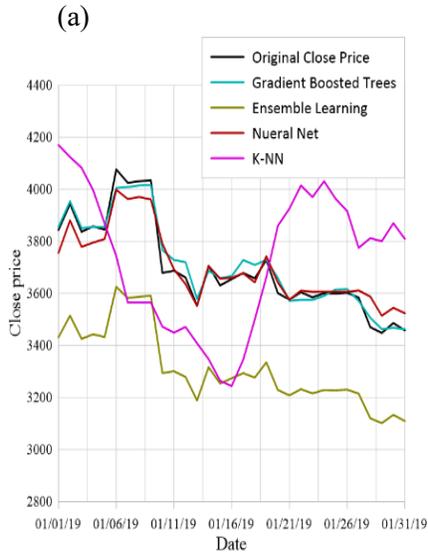

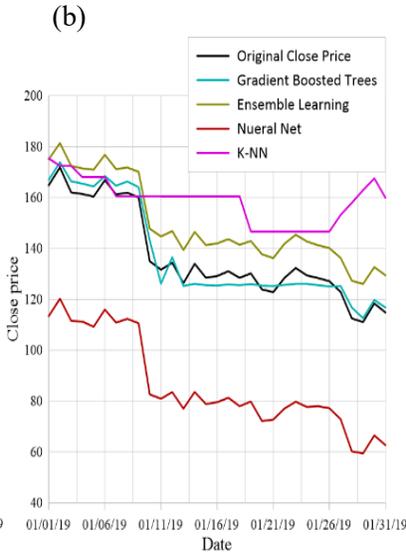

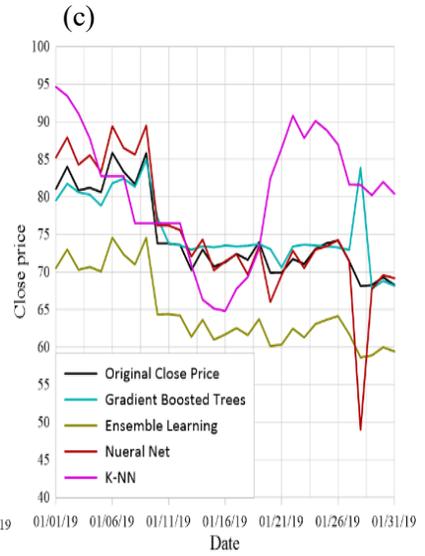

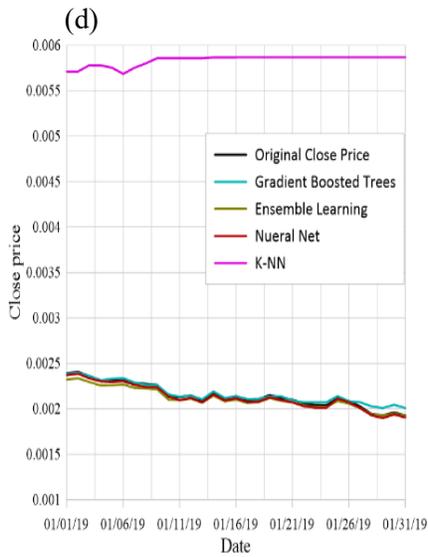

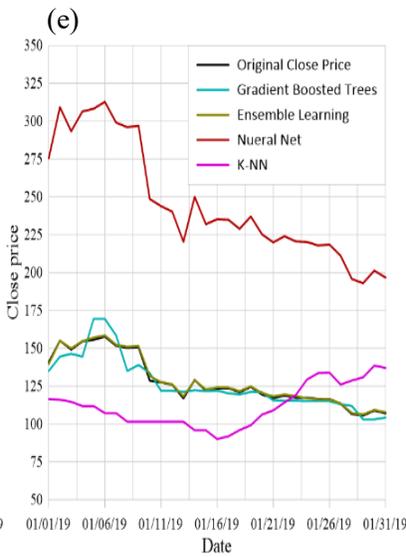

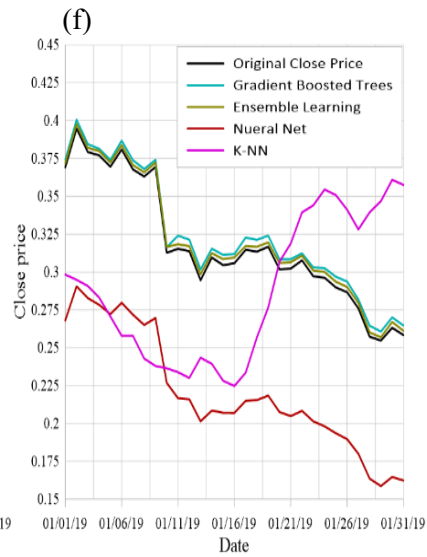

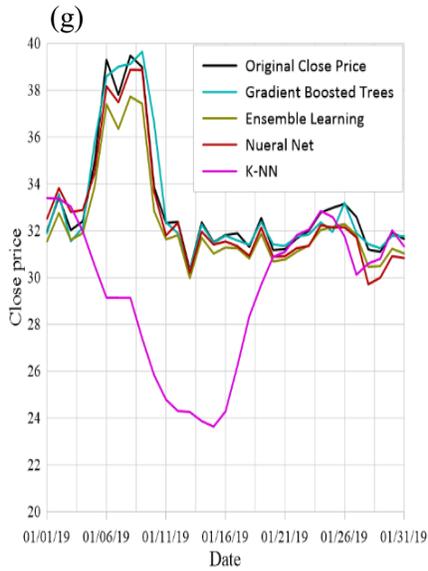

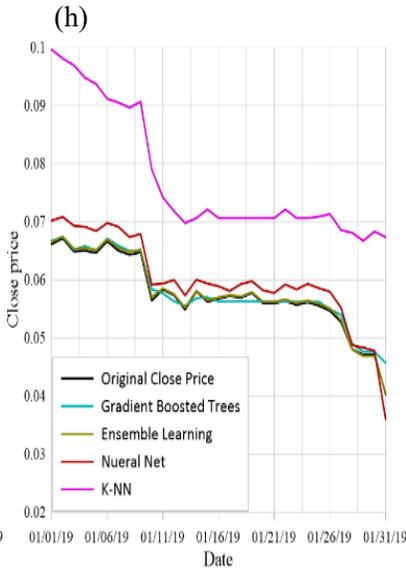

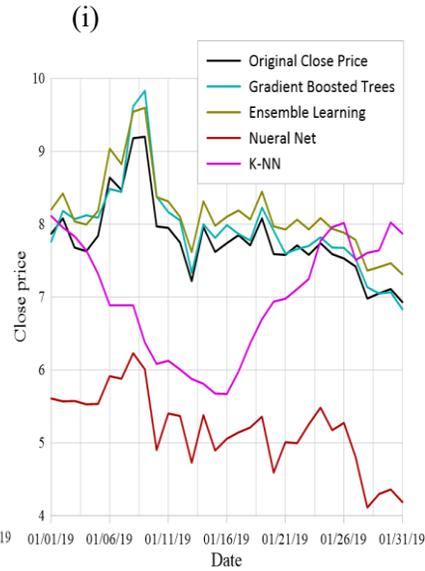

(j)

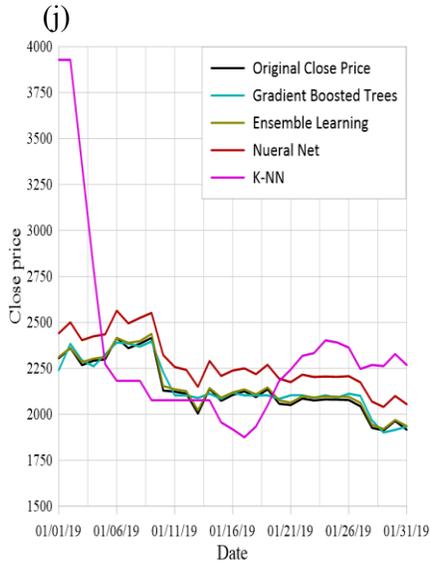

Figure 10. Comparison between original and forecasted close price obtained from RapidMiner using all models and methods of the month January 2019 (a) of constituent Bitcoin. (b) of constituent Bitcoin Cash. (c) of constituent Dash. (d) of constituent Dogecoin (DOGE). (e) of constituent Ethereum. (f) of constituent IOTA (MIOTA). (g) of constituent Litecoin. (h) of constituent NEM. (i) of constituent NEO. (j) of index (cci30).

# Supplement Information of "Predicting and Forecasting the Price of Constituents and Index of Cryptocurrency Using Machine Learning"

Reaz Chowdhury[1], M. Arifur Rahman[2], M. Sohel Rahman[3], M.R.C. Mahdy[1,4*]

[1]*Department of Electrical & Computer Engineering, North South University, Bashundhara, Dhaka 1229, Bangladesh*

[2]*Department of Accounting & Finance, North South University, Bashundhara, Dhaka 1229, Bangladesh*

[3]*Department of Computer Science & Engineering, Bangladesh University of Engineering & Technology, West Palasi, Dhaka 1205, Bangladesh*

[4]*Pi Labs Bangladesh LTD, ARA Bhaban,39, Kazi Nazrul Islam Avenue, Kawran Bazar, Dhaka 1215, Bangladesh*

*Corresponding Author: mahdy.chowdhury@northsouth.edu

**S1: Performance vector of Gradient Boosted Trees model**

| Predictive Model | Names of Constituents | Performance vector |
|---|---|---|
| Gradient Boosted Trees | Bitcoin Cash | Root mean squared error: 3.849 +/- 0.000<br>Prediction trend accuracy: 0.667<br>Absolute error: 3.429 +/- 1.748<br>Relative error: 2.52% +/- 1.30%<br>Squared error: 14.813 +/- 15.722<br>Correlation: 0.983<br>Squared correlation: 0.966 |
| | Bitcoin | Root mean squared error: 32.863 +/- 0.000<br>Prediction trend accuracy: 0.800<br>Absolute error: 25.005 +/- 21.324<br>Relative error: 0.68% +/- 0.57%<br>Squared error: 1079.974 +/- 1735.034<br>Correlation: 0.985<br>Squared correlation: 0.970 |
| | Dash | Root mean squared error: 3.297 +/- 0.000<br>Prediction trend accuracy: 0.733<br>Absolute error: 1.798 +/- 2.764<br>Relative error: 2.47% +/- 4.03%<br>Squared error: 10.872 +/- 43.503<br>Correlation: 0.809<br>Squared correlation: 0.655 |
| | Doge Coin (DOGE) | Root mean squared error: 0.000 +/- 0.000<br>Prediction trend accuracy: 0.800<br>Absolute error: 0.000 +/- 0.000<br>Relative error: 1.07% +/- 1.35%<br>Squared error: 0.000 +/- 0.000<br>Correlation: 0.986<br>Squared correlation: 0.971 |
| | Ethereum | Root mean squared error: 6.394 +/- 0.000<br>Prediction trend accuracy: 0.767<br>Absolute error: 4.980 +/- 4.010<br>Relative error: 3.66% +/- 2.55%<br>Squared error: 40.886 +/- 59.456<br>Correlation: 0.934<br>Squared correlation: 0.873 |
| | IOTA (MIOTA) | Root mean squared error: 0.006 +/- 0.000<br>Prediction trend accuracy: 0.667<br>Absolute error: 0.006 +/- 0.001<br>Relative error: 1.99% +/- 0.51%<br>Squared error: 0.000 +/- 0.000<br>Correlation: 1.000<br>Squared correlation: 0.999 |
| | Litecoin | Root mean squared error: 0.657 +/- 0.000<br>Prediction trend accuracy: 0.833 |

| | | Absolute error: 0.393 +/- 0.527 |
| | | Relative error: 1.15% +/- 1.53% |
| | | Squared error: 0.432 +/- 1.340 |
| | | Correlation: 0.969 |
| | | Squared correlation: 0.939 |
| | NEM | Root mean squared error: 0.001 +/- 0.000 |
| | | Prediction trend accuracy: 0.900 |
| | | Absolute error: 0.001 +/- 0.001 |
| | | Relative error: 1.56% +/- 2.27% |
| | | Squared error: 0.000 +/- 0.000 |
| | | Correlation: 0.984 |
| | | Squared correlation: 0.967 |
| | NEO | Root mean squared error: 0.237 +/- 0.000 |
| | | Prediction trend accuracy: 0.833 |
| | | Absolute error: 0.180 +/- 0.154 |
| | | Relative error: 2.26% +/- 1.84% |
| | | Squared error: 0.056 +/- 0.088 |
| | | Correlation: 0.961 |
| | | Squared correlation: 0.923 |

Figure 1s Performance vector values obtained from gradient boosted trees model using RapidMiner.

**S2: A sample of tree leaf obtained from gradient boosted trees model**

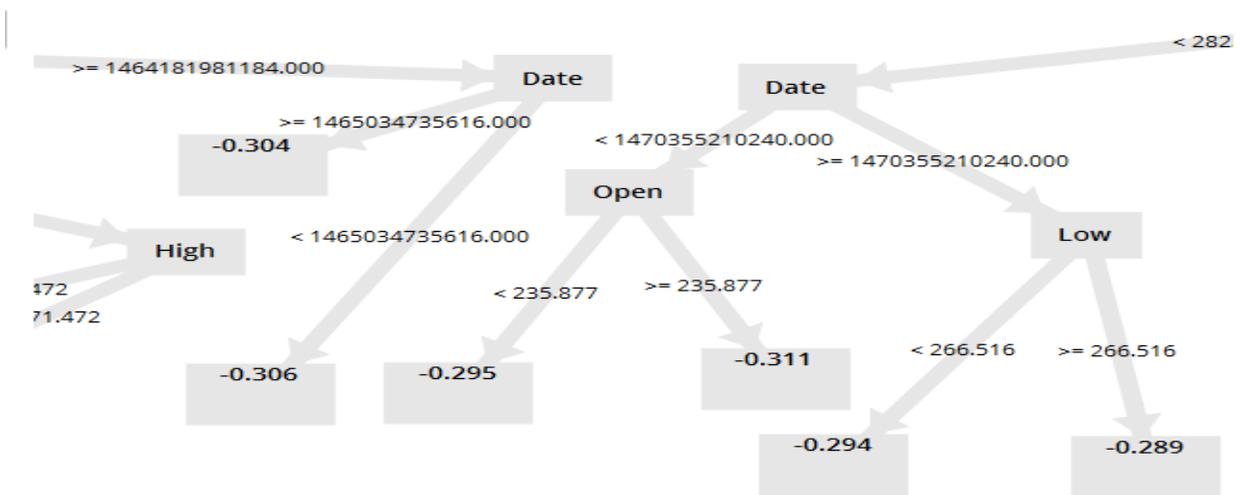

Figure 2s A sample of tree leaf obtained from model gradient boosted trees in RapidMiner.

## S3: Performance vector of neural network model

| Predictive Model | Names of Constituents | Performance vector |
|---|---|---|
| Neural Net | Bitcoin Cash | Prediction trend accuracy: 0.496 +/- 0.290 (micro average: 0.496)<br>Root mean squared error: 147.471 +/- 193.041 (micro average: 242.775 +/- 0.000)<br>Absolute error: 132.631 +/- 178.564 (micro average: 132.631 +/- 203.343)<br>Relative error: 13.95% +/- 11.66% (micro average: 13.95% +/- 14.16%)<br>Squared error: 58939.513 +/- 199131.986 (micro average: 58939.513 +/- 248098.729)<br>Correlation: 0.236 +/- 0.311 (micro average: 0.931)<br>Squared correlation: 0.152 +/- 0.254 (micro average: 0.866) |
| | Bitcoin | Prediction trend accuracy: 0.488 +/- 0.284 (micro average: 0.488)<br>Root mean squared error: 222.789 +/- 497.935 (micro average: 545.379 +/- 0.000)<br>Absolute error: 199.332 +/- 454.351 (micro average: 199.332 +/- 507.647)<br>Relative error: 5.91% +/- 5.81% (micro average: 5.91% +/- 6.84%)<br>Squared error: 297438.474 +/- 1489764.423 (micro average: 297438.474 +/- 1876854.100)<br>Correlation: 0.262 +/- 0.333 (micro average: 0.988)<br>Squared correlation: 0.179 +/- 0.283 (micro average: 0.976) |
| | Dash | Prediction trend accuracy: 0.483 +/- 0.282 (micro average: 0.483)<br>Root mean squared error: 16.964 +/- 45.259 (micro average: 48.322 +/- 0.000)<br>Absolute error: 15.350 +/- 42.210 (micro average: 15.350 +/- 45.819)<br>Relative error: 10.77% +/- 11.41% (micro average: 10.77% +/- 13.25%)<br>Squared error: 2334.972 +/- 13797.908 (micro average: 2334.972 +/- 15802.119)<br>Correlation: 0.242 +/- 0.319 (micro average: 0.977) |

| | | Squared correlation: 0.161 +/- 0.264 (micro average: 0.954) |
|---|---|---|
| | Doge Coin (DOGE) | Prediction trend accuracy: 0.473 +/- 0.282 (micro average: 0.473) |
| | | Root mean squared error: 0.000 +/- 0.000 (micro average: 0.001 +/- 0.000) |
| | | Absolute error: 0.000 +/- 0.000 (micro average: 0.000 +/- 0.001) |
| | | Relative error: 9.69% +/- 10.37% (micro average: 9.69% +/- 11.98%) |
| | | Squared error: 0.000 +/- 0.000 (micro average: 0.000 +/- 0.000) |
| | | Correlation: 0.246 +/- 0.328 (micro average: 0.965) |
| | | Squared correlation: 0.168 +/- 0.276 (micro average: 0.932) |
| | Ethereum | Prediction trend accuracy: 0.503 +/- 0.279 (micro average: 0.503) |
| | | Root mean squared error: 23.649 +/- 43.492 (micro average: 49.491 +/- 0.000) |
| | | Absolute error: 21.193 +/- 40.116 (micro average: 21.193 +/- 44.723) |
| | | Relative error: 10.30% +/- 9.87% (micro average: 10.30% +/- 11.42%) |
| | | Squared error: 2449.310 +/- 9480.333 (micro average: 2449.310 +/- 11688.171) |
| | | Correlation: 0.262 +/- 0.330 (micro average: 0.984) |
| | | Squared correlation: 0.177 +/- 0.275 (micro average: 0.968) |
| | IOTA (MIOTA) | Prediction trend accuracy: 0.511 +/- 0.277 (micro average: 0.511) |
| | | Root mean squared error: 0.197 +/- 0.306 (micro average: 0.364 +/- 0.000) |
| | | Absolute error: 0.178 +/- 0.283 (micro average: 0.178 +/- 0.318) |
| | | Relative error: 14.20% +/- 11.96% (micro average: 14.20% +/- 14.33%) |
| | | Squared error: 0.133 +/- 0.618 (micro average: 0.133 +/- 0.763) |
| | | Correlation: 0.236 +/- 0.319 (micro average: 0.938) |
| | | Squared correlation: 0.157 +/- 0.265 (micro average: 0.880) |
| | Litecoin | Prediction trend accuracy: 0.513 +/- 0.279 (micro average: 0.513) |
| | | Root mean squared error: 4.412 +/- 12.304 (micro average: 13.068 +/- 0.000) |
| | | Absolute error: 3.964 +/- 11.372 (micro average: 3.964 +/- 12.452) |
| | | Relative error: 8.17% +/- 8.84% (micro average: 8.17% +/- 10.20%)<br>Squared error: 170.765 +/- 1483.257 (micro average: 170.765 +/- 1683.792)<br>Correlation: 0.258 +/- 0.323 (micro average: 0.974)<br>Squared correlation: 0.171 +/- 0.269 (micro average: 0.948) |
| | NEM | Prediction trend accuracy: 0.503 +/- 0.282 (micro average: 0.503)<br>Root mean squared error: 0.020 +/- 0.059 (micro average: 0.063 +/- 0.000)<br>Absolute error: 0.018 +/- 0.054 (micro average: 0.018 +/- 0.060)<br>Relative error: 11.66% +/- 10.13% (micro average: 11.66% +/- 12.20%)<br>Squared error: 0.004 +/- 0.028 (micro average: 0.004 +/- 0.035)<br>Correlation: 0.259 +/- 0.326 (micro average: 0.965)<br>Squared correlation: 0.173 +/- 0.268 (micro average: 0.931) |
| | NEO | Prediction trend accuracy: 0.505 +/- 0.281 (micro average: 0.505)<br>Root mean squared error: 4.664 +/- 7.668 (micro average: 8.971 +/- 0.000)<br>Absolute error: 4.191 +/- 7.084 (micro average: 4.191 +/- 7.932)<br>Relative error: 13.67% +/- 15.30% (micro average: 13.67% +/- 17.05%)<br>Squared error: 80.482 +/- 337.773 (micro average: 80.482 +/- 405.483)<br>Correlation: 0.283 +/- 0.341 (micro average: 0.970)<br>Squared correlation: 0.196 +/- 0.289 (micro average: 0.940) |

Figure 3s Performance vector values obtained from neural network model using RapidMiner.

## S4: Performance vector of ensemble learning method

| Predictive method | Names of Constituents | Performance Vectors |
|---|---|---|

| Ensemble learner | Bitcoin Cash | Root mean squared error: 7.033 +/- 0.000<br>Prediction trend accuracy: 0.795<br>Absolute error: 4.771 +/- 5.167<br>Relative error: 3.17% +/- 4.93%<br>Squared error: 49.462 +/- 78.674<br>Correlation: 1.000<br>Squared correlation: 1.000 |
| --- | --- | --- |
| | Bitcoin | Root mean squared error: 963.826 +/- 0.000<br>Prediction trend accuracy: 0.516<br>Absolute error: 747.783 +/- 608.096<br>Relative error: 10.51% +/- 4.60%<br>Squared error: 928960.241 +/- 1391850.156<br>Correlation: 1.000<br>Squared correlation: 1.000 |
| | Dash | Root mean squared error: 63.579 +/- 0.000<br>Prediction trend accuracy: 0.555<br>Absolute error: 46.326 +/- 43.545<br>Relative error: 14.42% +/- 2.63%<br>Squared error: 4042.300 +/- 8038.797<br>Correlation: 1.000<br>Squared correlation: 1.000 |
| | Doge Coin (DOGE) | Root mean squared error: 0.000 +/- 0.000<br>Prediction trend accuracy: 0.755<br>Absolute error: 0.000 +/- 0.000<br>Relative error: 4.73% +/- 4.13%<br>Squared error: 0.000 +/- 0.000<br>Correlation: 0.999<br>Squared correlation: 0.998 |
| | Ethereum | Root mean squared error: 34.238 +/- 0.000<br>Prediction trend accuracy: 0.835<br>Absolute error: 18.565 +/- 28.769<br>Relative error: 2.50% +/- 2.79%<br>Squared error: 1172.272 +/- 2945.911<br>Correlation: 0.999<br>Squared correlation: 0.999 |
| | IOTA (MIOTA) | Root mean squared error: 0.002 +/- 0.000<br>Prediction trend accuracy: 0.924<br>Absolute error: 0.002 +/- 0.001<br>Relative error: 0.52% +/- 0.39%<br>Squared error: 0.000 +/- 0.000<br>Correlation: 1.000<br>Squared correlation: 1.000 |

| | Litecoin | Root mean squared error: 13.708 +/- 0.000 <br> Prediction trend accuracy: 0.665 <br> Absolute error: 8.547 +/- 10.717 <br> Relative error: 7.56% +/- 4.67% <br> Squared error: 187.909 +/- 404.995 <br> Correlation: 1.000 <br> Squared correlation: 1.000 |
|---|---|---|
| | NEM | Root mean squared error: 0.049 +/- 0.000 <br> Prediction trend accuracy: 0.866 <br> Absolute error: 0.019 +/- 0.045 <br> Relative error: 2.75% +/- 4.14% <br> Squared error: 0.002 +/- 0.009 <br> Correlation: 0.999 <br> Squared correlation: 0.998 |
| | NEO | Root mean squared error: 0.445 +/- 0.000 <br> Prediction trend accuracy: 0.816 <br> Absolute error: 0.383 +/- 0.227 <br> Relative error: 2.24% +/- 1.80% <br> Squared error: 0.198 +/- 0.259 <br> Correlation: 1.000 <br> Squared correlation: 1.000 |

Figure 4s Performance vector values obtained from ensemble learning method using RapidMiner.

## S5: Performance vector of K-NN model

| Forecasting Model | Names of Constituents | Performance Vectors |
|---|---|---|
| K-NN | Bitcoin | Root mean squared error: 2044.436 +/- 0.000 <br> Prediction trend accuracy: 0.452 <br> Absolute error: 856.854 +/- 1856.212 <br> Relative error: 23.82% +/- 22.89% <br> Squared error: 4179719.409 +/- 16760517.654 <br> Correlation: 0.859 <br> Squared correlation: 0.737 |
| | Bitcoin Cash | Root mean squared error: 939.920 +/- 0.000 <br> Prediction trend accuracy: 0.409 <br> Absolute error: 667.710 +/- 661.524 |

| | | Relative error: 82.32% +/- 92.99%<br>Squared error: 883450.416 +/- 1562483.485<br>Correlation: 0.105<br>Squared correlation: 0.011 |
|---|---|---|
| | Dash | Root mean squared error: 190.460 +/- 0.000<br>Prediction trend accuracy: 0.464<br>Absolute error: 75.325 +/- 174.932<br>Relative error: 46.76% +/- 57.06%<br>Squared error: 36275.091 +/- 133446.804<br>Correlation: 0.691<br>Squared correlation: 0.478 |
| | Doge Coin (DOGE) | Root mean squared error: 0.002 +/- 0.000<br>Prediction trend accuracy: 0.439<br>Absolute error: 0.001 +/- 0.002<br>Relative error: 40.55% +/- 35.09%<br>Squared error: 0.000 +/- 0.000<br>Correlation: 0.597<br>Squared correlation: 0.356 |
| | Ethereum | Root mean squared error: 213.076 +/- 0.000<br>Prediction trend accuracy: 0.459<br>Absolute error: 114.013 +/- 180.006<br>Relative error: 46.21% +/- 35.70%<br>Squared error: 45401.204 +/- 118636.508<br>Correlation: 0.783<br>Squared correlation: 0.613 |
| | IOTA (MIOTA) | Root mean squared error: 1.357 +/- 0.000<br>Prediction trend accuracy: 0.436<br>Absolute error: 0.880 +/- 1.033<br>Relative error: 74.62% +/- 72.87%<br>Squared error: 1.843 +/- 3.564<br>Correlation: 0.113<br>Squared correlation: 0.013 |
| | Litecoin | Root mean squared error: 37.930 +/- 0.000<br>Prediction trend accuracy: 0.462<br>Absolute error: 14.136 +/- 35.198<br>Relative error: 28.66% +/- 27.69%<br>Squared error: 1438.722 +/- 6901.387<br>Correlation: 0.768<br>Squared correlation: 0.590 |
| | NEM | Root mean squared error: 0.325 +/- 0.000<br>Prediction trend accuracy: 0.463 |

| | | Absolute error: 0.131 +/- 0.297<br>Relative error: 57.80% +/- 74.45%<br>Squared error: 0.106 +/- 0.355<br>Correlation: 0.409<br>Squared correlation: 0.168 |
|---|---|---|
| | NEO | Root mean squared error: 30.856 +/- 0.000<br>Prediction trend accuracy: 0.465<br>Absolute error: 17.882 +/- 25.146<br>Relative error: 52.02% +/- 43.62%<br>Squared error: 952.083 +/- 2172.463<br>Correlation: 0.707<br>Squared correlation: 0.500 |

Figure 5s Performance vector values obtained from K-NN model using RapidMiner.

## S6: Performance vector obtained for predicting and forecasting cci30

| Name of Index | Names of Constituents | Performance Vectors |
|---|---|---|
| Cryptocurrencies index 30 (cci30) | Gradient Boosted Trees | Root mean squared error: 37.269 +/- 0.000<br>Prediction trend accuracy: 0.767<br>Absolute error: 30.362 +/- 21.612<br>Relative error: 1.43% +/- 1.05%<br>Squared error: 1388.952 +/- 2171.964<br>Correlation: 0.968<br>Squared correlation: 0.938 |
| | Neural Net | Prediction trend accuracy: 0.490 +/- 0.292 (micro average: 0.490)<br>Root mean squared error: 258.784 +/- 575.383 (micro average: 630.720 +/- 0.000)<br>Absolute error: 233.303 +/- 533.608 (micro average: 233.303 +/- 585.984)<br>Relative error: 6.02% +/- 5.92% (micro average: 6.02% +/- 6.98%)<br>Squared error: 397807.137 +/- 2043368.049 (micro average: 397807.137 +/- 2350603.492)<br>correlation: 0.254 +/- 0.328 (micro average: 0.986)<br>Squared correlation: 0.172 +/- 0.275 (micro average: 0.971 |
| | Ensemble | Root mean squared error: 639.452 +/- 0.000<br>Prediction trend accuracy: 0.713<br>Absolute error: 366.350 +/- 524.106<br>Relative error: 4.00% +/- 3.74% |

| | | Squared error: 408899.489 +/- 992371.037 |
| --- | --- | --- |
| | | Correlation: 1.000 |
| | | Squared correlation: 0.999 |
| | K-NN | Root mean squared error: 2915.663 +/- 0.000 |
| | | Prediction trend accuracy: 0.423 |
| | | Absolute error: 1304.753 +/- 2607.433 |
| | | Relative error: 30.34% +/- 29.80% |
| | | Squared error: 8501088.395 +/- 28794869.536 |
| | | Correlation: 0.749 |
| | | Squared correlation: 0.561 |

Figure 6s Performance vector values obtained using gradient boosted trees, neural net, ensemble learning method and K-NN models of cryptocurrency index 30 (cci30).